\documentclass[a4paper,final,10pt]{article}
\usepackage[utf8]{inputenc}

\usepackage[T1]{fontenc}
\usepackage[french]{babel} 
\usepackage{amsmath}
\usepackage{txfonts}
\usepackage{enumitem}
\usepackage{geometry} 
\usepackage{fancyhdr, eso-pic}
\usepackage{xcolor,graphicx} 
\usepackage{hyperref}
\usepackage{multirow}   
\usepackage{listings}   
\usepackage{tikz}

\usepackage{siunitx}    
\makeatletter
    \ifdefined\qty
    \else
      \newcommand{\qty}[1]{\SI{#1}}
    \fi
\makeatother

\hypersetup{colorlinks=true, linkcolor=blue, filecolor=red, urlcolor=red, citecolor=teal}
\setlist[itemize,1]{label=\textbullet, partopsep=0pt, itemsep=3pt, topsep=6pt, parsep=0pt}

\lstnewenvironment{code}[1][]{\lstset{#1}}{} 
\makeatletter
\def\ifdraft{\ifdim\overfullrule>\z@
  \expandafter\@firstoftwo\else\expandafter\@secondoftwo\fi}
\renewcommand*{\maketitle}{
   \AddToShipoutPicture*{
    \put(0,0){\parbox[b][\paperheight]{\paperwidth}{%
        \includegraphics[width=\paperwidth,height=\paperheight]{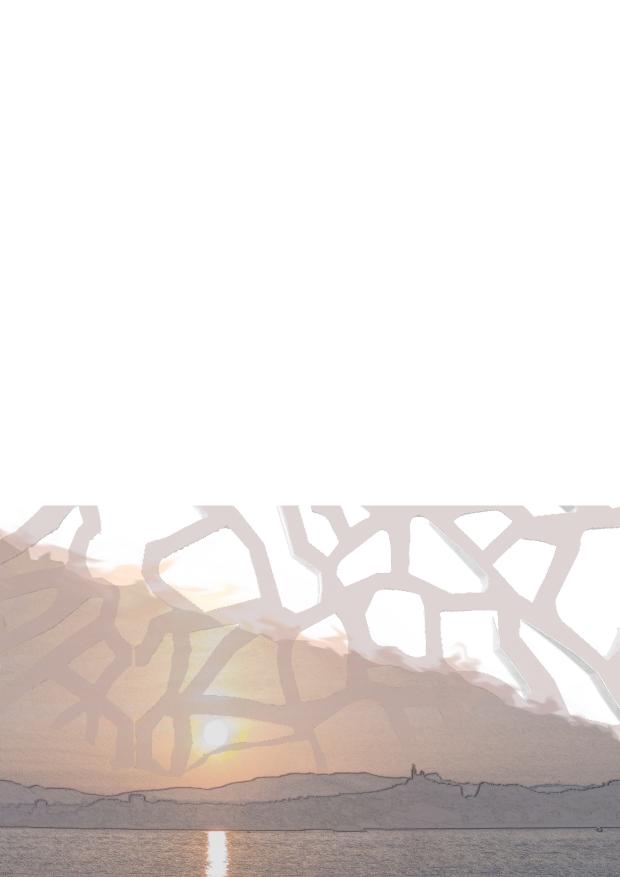}
    }}}
    \begin{titlepage}
    \thispagestyle{empty}
    \centering
    \setlength{\fboxrule}{2pt}
    \parbox[c][.99\textheight][t]{.99\textwidth}{
    \begin{center}
    \includegraphics[width=55mm,keepaspectratio=true]{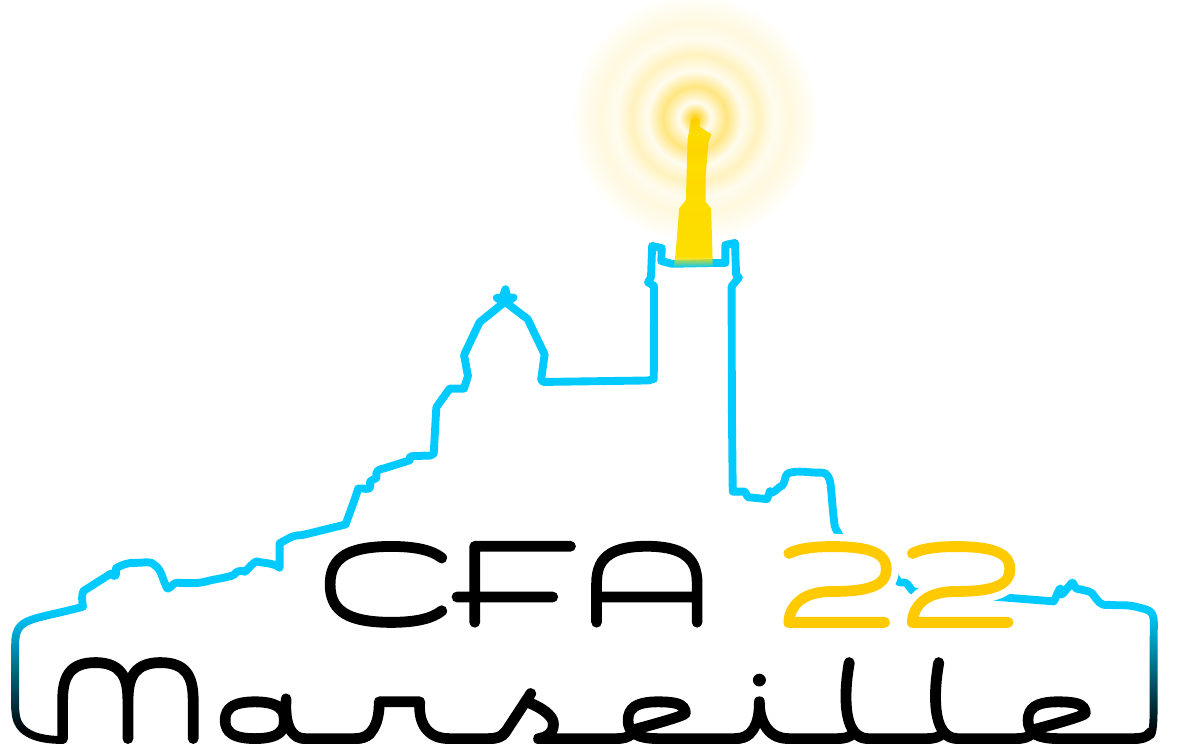}\\[.5cm]
    {\Large 16\textsuperscript{ème} Congrès Français d'Acoustique\unskip\strut\par}
    {\Large 11-15 Avril 2022, Marseille\unskip\strut\par}
  \vspace{1cm}
    {\LARGE \@title \par}%
    \vskip 3em%
    {\large
     \lineskip .75em%
      \begin{tabular}[t]{c}%
        \@author
      \end{tabular}\par}
    \end{center}}%
  \end{titlepage}
}
\renewcommand{\section}{\@startsection{section}{1}{0pt}{21pt}{10pt}{\Large\bfseries}}
\renewcommand{\subsection}{\@startsection{subsection}{2}{0pt}{10pt}{5pt}{\large\bfseries}}
\makeatother

\newcommand{\resume}[1]{%
\twocolumn[
  \begin{list}{}{%
    \setlength{\topsep}{0pt}%
    \setlength{\leftmargin}{1cm}%
    \setlength{\rightmargin}{1cm}%
    \setlength{\listparindent}{\parindent}%
    \setlength{\itemindent}{\parindent}%
    \setlength{\parsep}{\parskip}}%
    \item #1
  \end{list}\vspace{24pt}]
  }

\title{Design of an electronic circuit for loudspeaker real-time digital signal processing}
\author{
O.~Munroe,~ 
S.~Letourneur,~
A.~Novak
\\
{\small Laboratoire d'Acoustique de l'Université du Mans (LAUM), UMR 6613, }\\{\small Institut d'Acoustique - Graduate School (IA-GS), CNRS, Le Mans Université, France } 
\and 
}
\date{}  

\def\IIS{I$^2$S}

\begin{document}
\maketitle 

\resume{In modern audio systems, real-time digital signal processing algorithms are widely used for a variety of applications. The possibility of using a simple electronic circuit for variety of research projects has shown remarkable potential and is gradually attracting more and more attention from researchers and engineers. This contribution describes a design of such a board used in the framework of a PhD thesis whose subject is centred on the real-time correction of loudspeaker nonlinearities. The solution chosen in this work is based on a Teensy 3.6 microcontroller which is easy to program using the Arduino IDE and the libraries provided by Teensy. Two solutions are provided: one with an Audio board available on the market and another with a homemade board. Both solutions contain two inputs and at least one output (all 16 bits). This contribution does not detail the compensation algorithm related to the loudspeaker nonlinearities but focuses on the boards design, comparison of proposed solutions, and provides the basic codes to perform the real-time digital signal processing.}

\section{Introduction}
\par The work presented in this paper is motivated by the growing need to implement discrete real-time algorithms in research projects in universities and research laboratories. On the one hand, these projects are usually led by specialists in the field of acoustics or electro-acoustics who do not necessarily have expertise in electronics and in the design of processor boards. On the other hand, there are many solutions that are simple to use and require very little of this expertise. The purpose of this article is to present two inexpensive solutions that are currently being used in our laboratory for research projects to run our algorithms in real time.

\par Algorithms require a processor to perform the mathematical operations that must be performed quickly because the algorithms must be used in real time. They also need information from the analog domain: the stimulus and any other signals required, such as those used for feedback. These analog stimuli must therefore be conditioned and discretised using an ADC (analog-to-digital converter) for use in the processor, and then converted back into the analog domain using a DAC (digital-to-analog converter).

\par The choice of a processor and converters is usually based on the need for at least two 16-bit channels, the ability to perform very fast floating point calculations, low cost availability and ease of use. The Teensy boards meet all of these requirements while using the Arduino integrated development environment (IDE). The programs are coded in C and C++, using the free TeensyDuino add-on which contains numerous libraries.

\section{Teensy 3.6 Board}
\par
\begin{figure}[h]
    \centering
    \includegraphics[width=0.7\columnwidth,keepaspectratio=true]{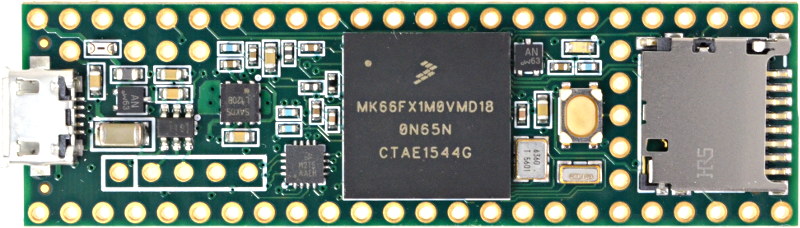}
    \caption{A picture of a Teensy 3.6 board.}
    \label{fig:teensy36}
\end{figure}

 The Teensy~3.6 (Fig.~\ref{fig:teensy36}) is based on an ARM Cortex M4 clocked at 180~MHz, which can be overclocked to 240~MHz in the IDE. It has a floating point unit (FPU) which is used to perform 32~bit hardware math operations on floating point numbers. It is available from various retailers for a price of about 35 euros.

\par The rest of the paper gives an overview of the use of the Teensy~3.6 card for real-time processing in a research project. Two solutions are proposed and compared in terms of latency, distortion and noise.

\begin{figure}[h]
    \centering
    \includegraphics[width=0.4\columnwidth,keepaspectratio=true]{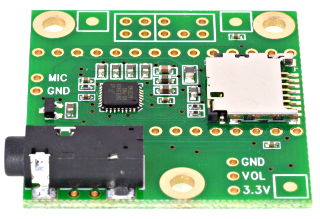}
    \caption{A picture of an Audio Adaptor Board for Teensy.}
    \label{fig:teensy_audio}
\end{figure}

\section{Teensy Board with \IIS}

\subsection{Hardware}

\par Teensy~3.6 includes hardware support for \IIS, a serial data protocol that handles high quality digital audio. Therefore, any circuit equipped with \IIS~can be connected to the Teensy board. In this work, we use the audio adapter board for Teensy (Fig.~\ref{fig:teensy_audio}). The board is equipped with an SGTL5000 chip with support for \IIS and is available on the market for the price of 20~EUR. The board includes a 2~channel line level input, a 2~channel line level output as pins on which connectors (BNC, jack, rca, ...) can be soldered. The card also includes a stereo headphone output and a mono microphone input.

\subsection{Software}

\par The Audio library that is part of the TeensyDuino installation contains classes such as {\tt AudioInputI2S} and {\tt AudioOutputI2S} that are very easy to use. In addition, it contains many built-in functions such as waveform generators, audio effects, fft analysis, etc. that can be directly used. Since research projects often require us to build our own algorithms, we provide in Appendix A the complete Teensyduino code with a class called {\tt audio\_processor} that provides direct access to input samples {\tt inL} and {\tt inR} and output samples {\tt outL} and {\tt outR} in type {\tt float}. The code can also be found on github \cite{TeensyI2S}.

\par The Audio library provides access to a buffer containing input and output samples. The default buffer length is 128 samples and can be modified in the library file {\tt AudioStream.h} which can be found in the Teensyduino installation folder in the folder  {\tt \char`\\hardware\char`\\teensy\char`\\avr\char`\\cores\char`\\teensy3} . The sampling frequency is set to 44.1~kHz, its value is defined in the same file.

\subsection{Characteristics}

\par To provide the main characteristics of the map, we focus on measuring three main limitations to system performance: latency, distortion and noise. All measurements are performed using Matlab and an RME~Fireface~400 sound card.

\par The latency of the system is a difference in time between the moment when a signal is introduced into the system and the moment when it appears at the output. We estimate the latency from the impulse response which is measured using a Maximum-Length Sequence (MLS) signal. The total harmonic distortion (THD) is then estimated from a 1~kHz sine wave with an amplitude of 0.5 Vrms as the ratio of the sum of the powers of all harmonic components to the power of the fundamental frequency. Finally, the THD+N (THD and noise) is estimated as the ratio between the power of the signal from which the fundamental frequency is removed and the power of the fundamental frequency.

\par
\begin{figure}[h]
    \centering
    \includegraphics[width=0.9\columnwidth,keepaspectratio=true]{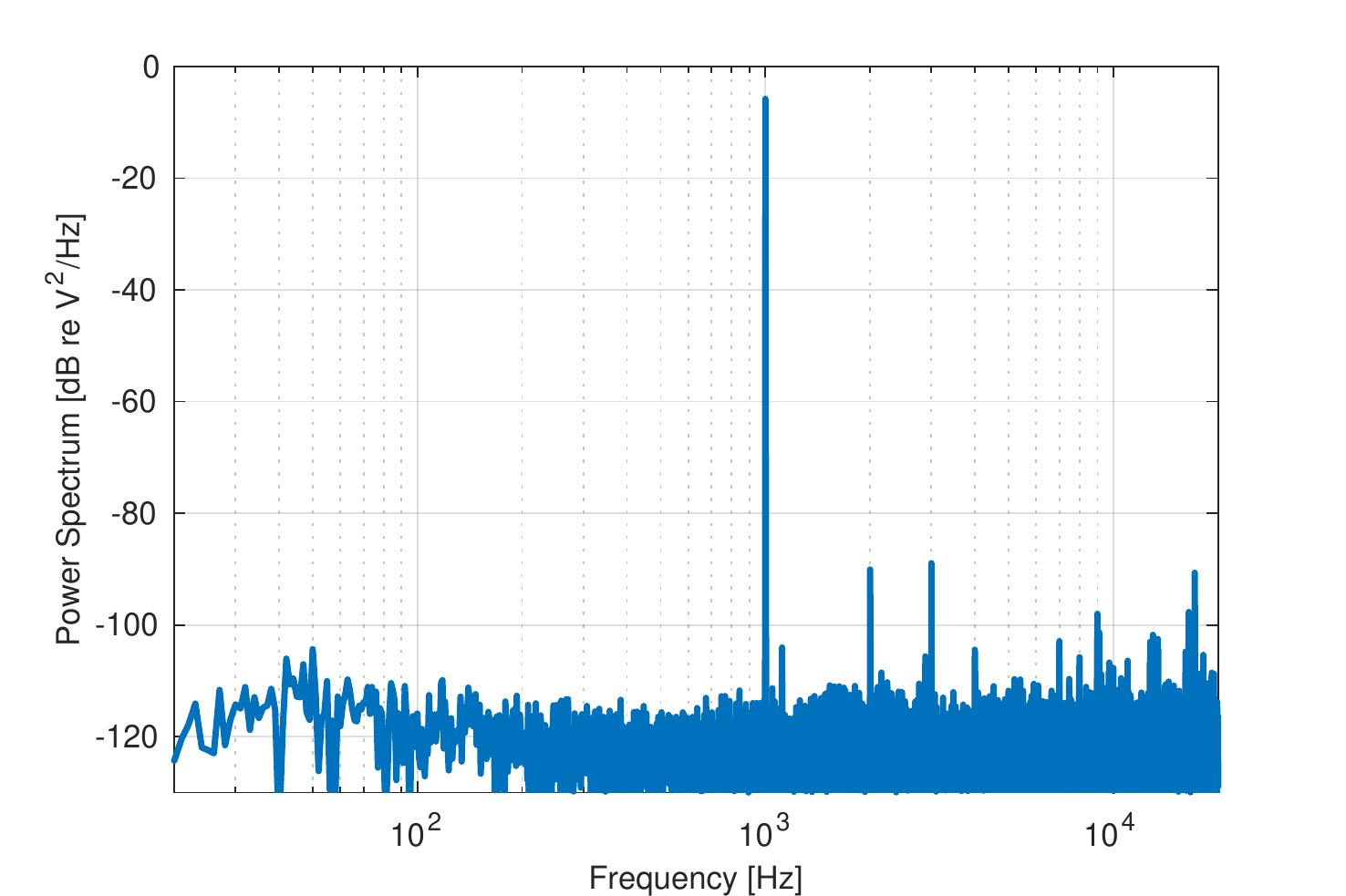}
    \caption{Output Power Spectrum of the Tennsy 3.6 with Audio Adapter (\IIS) excited with a 1~kHz sine wave with 0.5~Vrms amplitude.}
    \label{fig:result_I2S_THD_noise}
\end{figure}

 The latency results are compared below for several sizes of the parameters {\tt AUDIO\_BLOCK\_SAMPLES}. The THD and THD+N are independent of this parameter. Fig.~\ref{fig:result_I2S_THD_noise} then shows the output power spectrum when excited by a 1~kHz sine wave with an amplitude of 0.5~Vrms.

\renewcommand{\arraystretch}{1.5}
\begin{center}
\begin{tabular}{ c||c|c|c|c } 
{\tt AUDIO\_BLOCK\_SAMPLES} & 128    & 64     & 32   & 16 \\ 
 \hline
 Latency  [ms]                            & 9.24   & 4.9   & 2.7   & 1.63 \\ 
 THD                                 & \multicolumn{4}{c}{-80 dB \small \color{gray} @ 1 kHz, 0.5 Vrms} \\
 THD+N                             & \multicolumn{4}{c}{-68 dB \small \color{gray} @ 1 kHz, 0.5 Vrms} \\
\end{tabular}
\end{center}

\newpage
\section{Teensy Board with ADCs and external DAC}

\subsection{Hardware}
\par The other solution is to use the internal ADCs and external DACs. The Teensy~3.6 has two successive approximation (SAR) ADCs that can access 25 different pins on the Teensy PCB. The DACs on the board are only 12~bit, so an external DAC with a higher resolution is needed.

\subsubsection*{ADC and Signal Conditioning}
\par The ADCs can use an internal 3.3 V reference or an external reference voltage for conversion. The ADCs are capable of up to 16-bit resolution, but due to various sources of noise, the effective number of bits (ENOB) is typically specified at 13 bits for a single input. Thus, if overall noise is to be kept low, special care must be taken in the design of the signal conditioning section.

\par
\begin{figure}[h!]
    \centering
    \includegraphics[width=\columnwidth]{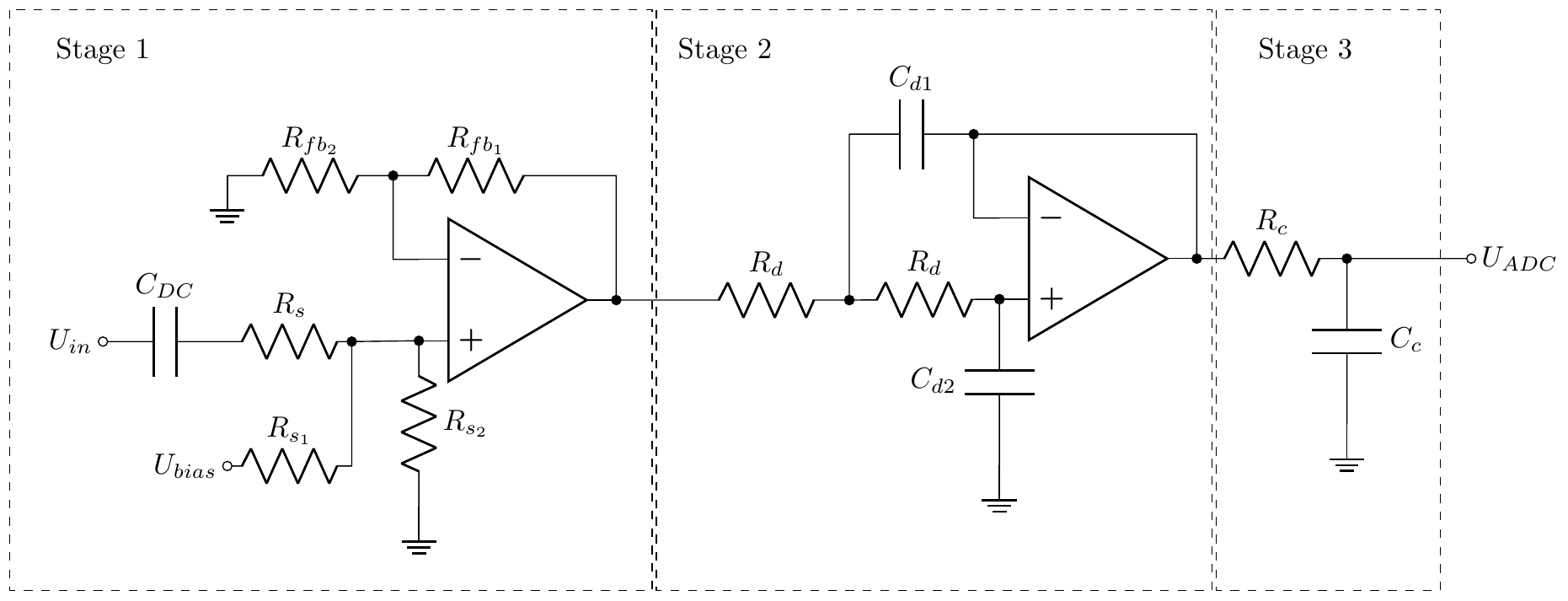}
    \caption{Analogue Input}
    \label{fig:AnalogIn}
\end{figure}

 To ensure accurate data with the least amount of noise, a proper analog signal conditioning circuit must be designed. The signal conditioning circuit consists of three stages (Fig.~\ref{fig:AnalogIn}). The first stage is where the signal is combined with the bias voltage required by the ADC. This bias voltage $U_{bias}$ is set to 1.65V, which is half the reference voltage of the ADC and is provided by a precision voltage reference, a MAX6043CAUT33+T from Maxim Integrated. To ensure that no external DC voltage hinders the performance of the $1^{st}$ stage, the $U_{in}$ input signal is AC-coupled via a $C_{DC}$ capacitor. The value of the capacitor is extremely large ($1~mF$) which ensures a very low cut-off frequency of about 40 mHz. The second stage is the one where the signal is low pass filtered by a Sallen Key filter of $2^{nd}$ order. The values of the components in this stage are chosen to define an anti-aliasing filter. It also reduces the amount of high frequency noise present at the ADC inputs. The final stage consists of two components, a resistor and a capacitor. The capacitor is there to provide a charge reserve for the SAR ADC while the resistor is there to help the stability of the operational amplifier (Op-Amp).

Due to the voltage limitations of the Teensy PCB, voltages above 3.5~V or below -0.2~V will damage the ADCs. There are several ways to protect the inputs, but in the interest of keeping the circuits simple and free of parasitic effects, the following solution was chosen. Rail to Rail operational amplifiers powered by a single regulated 3.3~V DC voltage were used in the signal conditioning stages. Thus, the outputs of the Op-Amps cannot be higher than the 3.3~V supply voltage. The chosen Op-Amp, the TI LME49721, can operate from a single supply up to 5.5~V and is capable of producing output voltages within 30~mV of the positive or negative supply. It is a low-noise low-distortion audio Op-Amp that is unity gain stable.

\subsubsection*{DAC}
The on-board DAC is only 12~bits and does not meet our needs. We therefore use an external 16~bit DAC, the PmodDA3, available on the market for 35~EUR. The PmodDA3 communicates with the Teensy board via an SPI (Serial Peripheral Interface) protocol. The 16-bit unsigned integer values of the data samples are first split into two 8-bit values. These values are then loaded into a buffer to be sent via SPI. The rest of the SPI transfer is configured as in the data sheet recommendations.

\begin{figure}[h!]
    \centering
    \includegraphics[width=0.4\columnwidth]{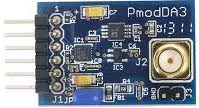}
    \caption{DAC PmodDA3 board.}
    \label{fig:PmodDA3}
\end{figure}

\subsection{Software}
Once the signal has reached the ADC, it is also a matter of ensuring a regular and precise timing of the conversion. This timing is provided by the programmable delay block (PDB). At a basic level, the PDB is set to a specific frequency, in this case the sample rate. The PDB is used to initiate a conversion in the ADCs. Once the conversion is complete, a flag is raised and the values are read and used in the algorithm. To get the samples from the ADCs, we use a free ADC library \cite{ADCLib} (provided by Pedro Villanueva) that makes setting up and using the ADCs incredibly simple. Full code for Teensyduino is available in the appendix and on github: \cite{TeensyADCDAC}.

\subsection{Characteristics}

\begin{figure}[h]
    \centering
    \includegraphics[width=0.9\columnwidth,keepaspectratio=true]{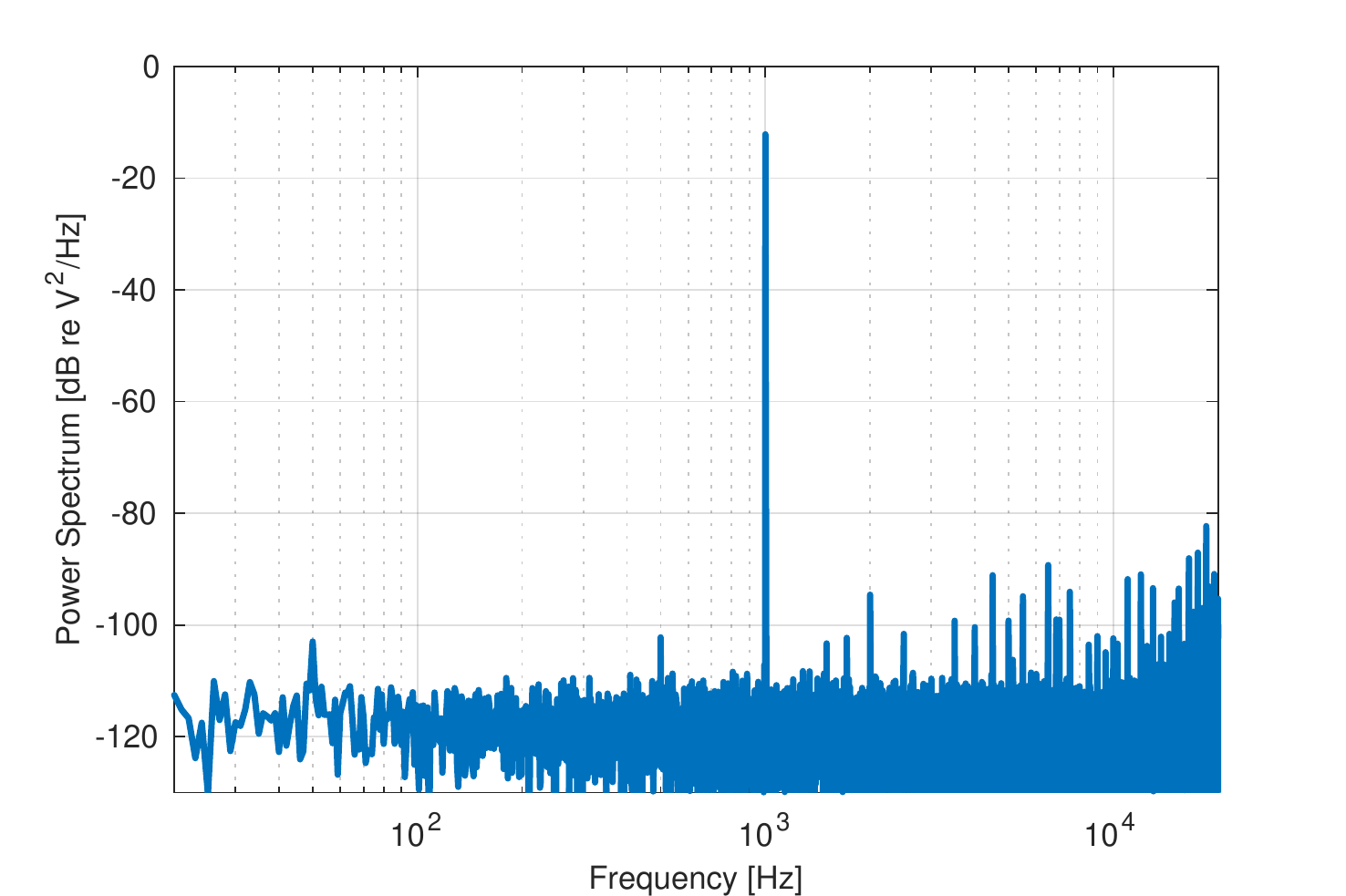}
    \caption{Output Power Spectrum of the Tennsy 3.6 with internal ADC and external DAC excited with a 1~kHz sine wave with 0.5~Vrms amplitude. Tha ADC sampling speed is set to {\tt LOW\_SPEED}.}
    \label{fig:result_DAC_THD_noise}
\end{figure}

This section summarizes the measured characteristics of the Teensy~3.6 card with internal ADCs and an external 16~bit DAC. All measurements provided in this section are for the 96~kHz sample rate. As the Pedvide ADC library allows to modify the ADC sampling speed, we test two configurations with different parameters of {\tt ADC\_SAMPLING\_SPEED} : {\tt LOW\_SPEED} and {\tt HIGH\_SPEED}. Fig.~\ref{fig:result_DAC_THD_noise}. The measured latency, THD and THD+N are summarized below.  Fig.~\ref{fig:result_DAC_THD_noise} then shows the output power spectrum when excited by a 1~kHz sine wave with an amplitude of 0.5~Vrms.

\renewcommand{\arraystretch}{1.5}
\begin{center}
\begin{tabular}{ c||c|c } 
{\tt ADC\_SAMPLING\_SPEED} & {\tt LOW\_SPEED}    & {\tt HIGH\_SPEED} \\ 
 \hline
 Latency  [$\mu$s]                             &  12  & 9.6 \\ 
 THD [dB]                                           &  -76 & -67 \\
 THD+N [dB]                                      &  -63 & -61 \\
\end{tabular}
\end{center}

\section{Conclusion}
In this paper we compare two solutions with the Teensy~3.6 board used for real-time audio signal processing in research projects. The solution based on the Audio Adaptor Board is much simpler to use than the solution with internal ADCs and external DACs for which an external board with components for signal conditioning was used. When comparing the THD, the Audio Adaptor Board shows lower distortion than the ADC/DAC solution. However, the latency of the ADC/DAC solution is about 10~$\mu$s (micro seconds) whereas the latency of the Audio Adaptor Board solution is several milliseconds depending on the buffer size. We therefore keep both solutions for real-time processing in our research and even student projects. When low latency is not required, we recommend using the Audio Adaptor Board for its simplicity. For projects requiring very low latency, we use the board with internal ADCs and external DACs.

\clearpage
\section*{Appendix A: codes for \IIS}
\label{sec:AppendixA}
\noindent \tt{\bf teensy\_i2s\_IO\_processor.ino}
\hrule
\begin{code}
#include <Audio.h>
#include "audio_processor.h"

// --- connections
AudioInputI2S i2s1;
AudioOutputI2S i2s2;

AudioProcessor proc; 

AudioConnection patchCord1(i2s1, 0, proc, 0);
AudioConnection patchCord2(i2s1, 1, proc, 1);
AudioConnection patchCord3(proc, 0, i2s2, 0);
AudioConnection patchCord4(proc, 1, i2s2, 1);

AudioControlSGTL5000 sgtl5000_1;

void setup() {
  AudioMemory(12);

  // Enable the audio shield with LineIn
  sgtl5000_1.enable();  
  sgtl5000_1.inputSelect(AUDIO_INPUT_LINEIN);

  // set volume level of lineIn
  sgtl5000_1.lineInLevel(0); 
  // set volume level of linOut
  sgtl5000_1.lineOutLevel(13); 
}

void loop() {
}
\end{code}
\vspace{1cm}

\noindent \tt{\bf audio\_processor.h}
\hrule
\begin{code}
#ifndef audio_processor_h_
#define audio_processor_h_

#include "AudioStream.h"

class AudioProcessor : public AudioStream
{
public:
  AudioProcessor(void) : AudioStream(2, inputQueueArray) {}
  virtual void update(void);

  const int resolutionDAC = 16;
  const int resolutionADC = 16;
  const float conversionADC = 1.0f/((1<<resolutionADC)-1);
  const float conversionDAC = (1<<resolutionDAC)-1;

private:
  audio_block_t *inputQueueArray[2];
};

#endif
\end{code}

\newpage
\noindent \tt{\bf audio\_processor.cpp}
\hrule
\begin{code}
#include "audio_processor.h"

void AudioProcessor::update(void)
{
  audio_block_t *blockL, *blockR;
  float inL, inR, outL, outR;
  unsigned int i;

  blockL = receiveWritable(0);
  if (!blockL) return;
  blockR = receiveWritable(1);
  if (!blockR) return;

  for (i=0; i < AUDIO_BLOCK_SAMPLES; i++) {
    // read the input signal
    inL = blockL->data[i] * conversionADC;
    inR = blockR->data[i] * conversionADC;

    // processing
    outL = inL;
    outR = inR;

    // write the output signal
    blockL->data[i] = outL * conversionDAC;
    blockR->data[i] = outR * conversionDAC;    
  }

  transmit(blockL,0);
  transmit(blockR,1);
  release(blockL);
  release(blockR);
}

\end{code}

\begin{tikzpicture}[remember picture, overlay]
    \fill[blue, opacity=0.05] (-0.6,2.6)  rectangle ++(9, 5.5);
    \draw[blue, line width=2] (0,4.5)  rectangle ++(7, 1.6);
\end{tikzpicture}

\clearpage
\onecolumn
\section*{Appendix B: codes for ADC and DAC}
\label{sec:AppendixB}
\begin{code}
#include <ADC.h>
#include <ADC_util.h>
#include <SPI.h>

//Input (ADC) initialisation
const int readPin1 = A0;
const int readPin2 = A2;
const float VrefADC = 3.3;
const int ADCAverages = 1;
volatile uint16_t valADC0 = 0, valADC1 = 0;
volatile uint8_t VALADC0Ready = false, VALADC1Ready = false;
volatile float input1 = 0.0, input2 = 0.0;
const int resolutionADC = 16;
const float conversionConstADC = VrefADC/((1<<resolutionADC)-1);


//Output (DAC) signal initialisation
const int slaveSelectPin = 25;
const int DIO = 2;
const int LDAC = 24;
const float VrefDAC = 2.5;
volatile uint16_t valDAC = 0;
volatile float val4DACOut = 0.0;
const int resolutionDAC = 16;
const float conversionConstDAC = ((1<<resolutionDAC)-1)/VrefDAC;

// Timing
const float sampleRateHz = 48000.0;
const int PmodDA3SPIMHz = 50;
const long PmodDA3SPIHz = PmodDA3SPIMHz*1000*1000;

// Using the Pedvide ADC Library on Github
ADC *adc = new ADC(); // adc object

void setup(void)
{
  // Declare needed pins as inputs or outputs
  pinMode(readPin1, INPUT);
  pinMode(readPin2, INPUT);
  pinMode (slaveSelectPin, OUTPUT);
  pinMode(DIO, OUTPUT);
  pinMode (LDAC, OUTPUT);

  // When using the Pedvide ADC library we can set more options
  adc->adc0->setAveraging(ADCAverages); // set number of averages
  adc->adc0->setResolution(resolutionADC); // set bits of resolution
  adc->adc0->setReference(ADC_REFERENCE::REF_3V3); // Set voltage reference for ADC.
  adc->adc0->setSamplingSpeed(ADC_SAMPLING_SPEED::LOW_SPEED); // change the sampling speed
  adc->adc0->setConversionSpeed(ADC_CONVERSION_SPEED::MED_SPEED); // change the conversion speed

  adc->adc0->stopPDB();
  adc->adc0->startSingleRead(readPin1);
  adc->adc0->enableInterrupts(adc0_isr);
  adc->adc0->startPDB(sampleRateHz); //frequency in Hz
  
  adc->adc1->setAveraging(ADCAverages); // set number of averages
  adc->adc1->setResolution(resolutionADC); // set bits of resolution
  adc->adc1->setReference(ADC_REFERENCE::REF_3V3); // Set voltage reference for ADC.
  adc->adc1->setSamplingSpeed(ADC_SAMPLING_SPEED::LOW_SPEED); // change the sampling speed
  adc->adc1->setConversionSpeed(ADC_CONVERSION_SPEED::MED_SPEED); // change the conversion speed

  adc->adc1->stopPDB();
  adc->adc1->startSingleRead(readPin2); 
  adc->adc1->enableInterrupts(adc1_isr);
  adc->adc1->startPDB(sampleRateHz); //frequency in Hz

  
  NVIC_SET_PRIORITY(IRQ_USBOTG, 200);
  // initialise the SPI channel
  SPI.begin();
  SPI.beginTransaction(SPISettings(PmodDA3SPIHz, MSBFIRST, SPI_MODE0));
}


void Operations(void)
{  
  // read ADC Value and convert to voltage
  input1 = (valADC0 * conversionConstADC - 1.625);
  input2 = (valADC1 * conversionConstADC - 1.625);

  // do the Signal Processing (example: mean value of both inputs)
  val4DACOut = input1*0.5 + input2*0.5;

  // convert the output value for DAC
  valDAC = ((val4DACOut+1.25)*conversionConstDAC);

  // send the value using the SPI
  uint8_t SPIBuff[2] = {0};
  SPIBuff[0] = valDAC >> 8;
  SPIBuff[1] = valDAC & 0xFF;
 
  digitalWrite(LDAC,HIGH); 
  // take the SS pin low to select the chip:
  digitalWrite(slaveSelectPin,LOW);
  //  send in the address and value via SPI:
  SPI.transfer(SPIBuff, 2);
  // take the SS pin high to de-select the chip:
  digitalWrite(slaveSelectPin,HIGH);
  digitalWrite(LDAC,LOW);
}

void loop(void)
{
  if (VALADC0Ready && VALADC1Ready){
    Operations();
    VALADC0Ready = false;
    VALADC1Ready = false;
  }
}

void adc0_isr() {
    valADC0 = (uint16_t)adc->adc0->readSingle();
    VALADC0Ready = true;
}

void adc1_isr() {
    valADC1 = (uint16_t)adc->adc1->readSingle();
    VALADC1Ready = true;
}
\end{code}

\begin{tikzpicture}[remember picture, overlay]
    \fill[blue, opacity=0.05] (-0.6,7.7)  rectangle ++(14, 10.5);
   \draw[blue, line width=2] (-0.3,14.6)  rectangle ++(13, 1.1);
\end{tikzpicture}


\small
\begin{thebibliography}{10}
	\bibitem{TeensyI2S} Novak, A. (2022). Teensy Audio Processor (I2S) (Version 1.0.0) [Computer software], \url{https://github.com/antonin-novak/teensy_audio_processor_I2S}
	
	\bibitem{TeensyADCDAC} Munroe, O., Letourneur, S., \& Novak, A. (2022). Teensy 16bit Audio Processor (ADC-DAC) (Version 1.0.0) [Computer software]
	\url{https://github.com/antonin-novak/teensy_16bit_processor_ADC_DAC}
	
	\bibitem{ADCLib} Pedro Villanueva (2020). Teensy ADC Library [computer Software]
	\url {http://pedvide.github.io/ADC/}
\end{thebibliography}
\end{document}